\documentclass[twocolumn]{article}
\usepackage{url}
\usepackage{amsmath}
\usepackage{authblk}
\usepackage{amssymb}
\usepackage{braket}
\usepackage{array}
\usepackage{tabularx}
\usepackage{ragged2e}
\usepackage{booktabs}
\usepackage{float}
\usepackage[top=1in, bottom=1in, left=0.8in, right=0.8in]{geometry}
\usepackage[table]{xcolor}

\title{An Overview of Josephson Junctions Based QPUs}

\usepackage{authblk}

\author[1,2]{Omid Mohebi}
\author[1]{Alireza Hesam Mohseni}

\affil[1]{Department of Computer Engineering, Faculty of Artificial Intelligence and Computing, University of Science and Culture, Tehran, Iran}
\affil[2]{National Brain Centre, Iran University of Medical Sciences, Tehran, Iran}

\date{\today} 
\usepackage{graphicx}

\providecommand{\keywords}[1]{\textbf{\textit{Keywords---}} #1}

\begin{document}

\maketitle

\begin{abstract}
Quantum processing units (QPUs) built on superconducting Josephson junctions remain the most industrially mature route to fault-tolerant quantum computing, but the field has moved substantially since early 2025. This paper provides an updated overview of Josephson-junction QPUs, grounded in the quantum-mechanical principles, superposition, entanglement, and decoherence, that any qubit implementation must contend with, and in the physics of Cooper pairing and quantum tunneling that make a Josephson junction behave as a controllable qubit. We examine the engineering challenges of scaling these devices, including crosstalk, classical control-interface bottlenecks, and quantum error correction, and discuss the first below-threshold surface-code demonstrations on superconducting hardware. We survey recent materials advances, near-term computational results, and applications beyond computing, and compare Josephson-junction QPUs against trapped-ion, photonic, and neutral-atom alternatives, the latter having emerged as a serious fourth architecture since our original analysis. We close with the current public roadmaps toward fault-tolerant superconducting quantum computers and a calibrated assessment of how much progress those roadmaps still require.
\end{abstract}

\\

\section{Introduction}
\label{sec:introduction}

The pursuit of computational power has driven technological progress for decades. Classical computers, built on bits that are strictly either 0 or 1, have served science and industry remarkably well, yet they reach fundamental limits when confronted with certain classes of problems: simulating molecular interactions at the quantum level, factoring large integers, or searching unstructured spaces of astronomical size. Quantum computers approach these problems differently, harnessing the principles of quantum mechanics rather than working around them.

At the core of a quantum computer is the quantum processing unit (QPU), which operates on qubits rather than bits. A qubit can exist in a superposition of the states $|0\rangle$ and $|1\rangle$ simultaneously, and multiple qubits can be entangled so that their combined state cannot be
described independently of one another. The two superposition and entanglement properties, give QPUs access to computational strategies with no classical counterpart, and they underlie the following broad application areas:

\begin{itemize}
    \item \textbf{Cryptography.} Shor's algorithm shows that a sufficiently large, fault-tolerant QPU could factor integers efficiently, threatening RSA-based cryptography while motivating quantum-resistant and quantum-native schemes such as quantum key distribution \cite{Shor,NielsenChuang}.
    \item \textbf{Quantum simulation.} QPUs can represent quantum systems natively, offering a route to simulating molecular and materials behavior with an accuracy that is exponentially costly to reach classically, with applications from drug discovery to aerospace materials \cite{CaoRomeroAspuru}.
    \item \textbf{Machine learning.} Quantum machine learning remains an early-stage field, but hybrid quantum-classical algorithms are actively explored as a way to accelerate specific learning and sampling tasks \cite{SchuldSinayskiyPetruccione}. \item \textbf{Optimization.} Variational and adiabatic quantum algorithms target combinatorial optimization problems in logistics, finance, and scheduling \cite{FarhiGoldstoneGutmann}.
\end{itemize}

Realizing these applications at scale requires physical qubits that can be built, controlled, and connected by the thousands to millions, while preserving quantum coherence long enough to run useful algorithms. Several physical platforms compete to satisfy these requirements, including trapped ions, photons, neutral atoms held in optical tweezers, and superconducting
circuits based on Josephson junctions. Each modality makes a different trade-off among coherence time, gate speed, connectivity, and manufacturability, and no single platform currently dominates on every axis; Section~\ref{sec:comparison} returns to this comparison in detail.

This paper focuses on the superconducting, Josephson-junction-based approach, which has attracted the largest combined industrial investment to date and, as of this writing, has produced the first experimental demonstrations of quantum error correction operating below the
surface-code threshold \cite{acharya2025willow}. We build the case for this platform from first principles. Section~\ref{sec:background} reviews the quantum-mechanical foundations, superposition, entanglement, decoherence, and the density-matrix formalism, that any qubit implementation must contend with. Section~\ref{sec:fundamentals} explains how Josephson
junctions exploit Cooper pairing and quantum tunneling to realize a controllable qubit. Section~\ref{sec:challenges} examines the engineering challenges of scaling these devices, including crosstalk, control complexity, and the overhead of quantum error correction. Section~\ref{sec:applications} surveys applications and materials-science advances, and Section~\ref{sec:comparison} compares Josephson-junction QPUs against ion-trap, photonic, and neutral-atom alternatives. Section~\ref{sec:outlook} discusses developments since our original 2025 analysis, and Section~\ref{sec:conclusion} concludes.

\section{Background: Quantum Mechanical Foundations}
\label{sec:background}

Before examining how Josephson junctions realize a physical qubit, we review the quantum-mechanical principles that any qubit implementation, superconducting or otherwise, must contend with: superposition, entanglement, decoherence, and the distinction between pure and mixed states. Readers already comfortable with these concepts may skip to Section~\ref{sec:fundamentals}.

\subsection{Superposition and Quantum Parallelism}

Superposition allows a quantum system to exist in a linear combination of basis states rather than a single definite state. A qubit's state is written as: 
\[
|\psi\rangle = \alpha|0\rangle + \beta|1\rangle, \qquad |\alpha|^2+|\beta|^2=1,
\]
where $\alpha$ and $\beta$ are complex probability amplitudes; $|\alpha|^2$ and $|\beta|^2$ give the probabilities of measuring $|0\rangle$ and $|1\rangle$, respectively \cite{NielsenChuang,TurroThesis}. A register of $n$ qubits can occupy a superposition of all $2^n$ basis states at once, and a quantum algorithm applied to that register acts on every basis state in parallel. This is the origin of \emph{quantum parallelism}: measurement still collapses the register to a single outcome, so the advantage lies not in reading out all $2^n$ results directly, but in algorithms engineered so that interference concentrates probability on the answer of interest. Grover's search algorithm and Shor's factoring algorithm are the canonical examples of this strategy translating into a provable speedup over the best known classical methods \cite{Grover1996,Shor1994, Shor}.

\subsection{Entanglement}

Entanglement is a correlation between two or more qubits that cannot be decomposed into independent single-qubit states. The four Bell states are the maximally entangled two-qubit states:
\[
|\Phi^{\pm}\rangle = \tfrac{1}{\sqrt{2}}(|00\rangle \pm |11\rangle), \qquad
|\Psi^{\pm}\rangle = \tfrac{1}{\sqrt{2}}(|01\rangle \pm |10\rangle).
\]
Measuring one qubit of an entangled pair instantaneously determines the outcome of measuring the other, a correlation stronger than any classical hidden-variable model can reproduce. The 1935 Einstein-Podolsky-Rosen argument \cite{EinsteinPodolskyRosen1935} questioned whether this implied quantum mechanics was incomplete; Bell's theorem \cite{Bell1964} and its subsequent experimental tests \cite{Clauser1969} instead confirmed that entanglement is a genuine, non-classical resource. It underlies quantum teleportation, quantum key distribution, superdense coding, and, critically for this paper, quantum error correction, which encodes a single logical qubit across several physical qubits held in an entangled state \cite{NielsenChuang,TurroThesis}. Entanglement is also fragile: because it depends on precise phase relationships between qubits, it is highly susceptible to the environmental interactions discussed next.

\begin{figure*}[t]
    \centering
    \includegraphics[width=0.5\textwidth]{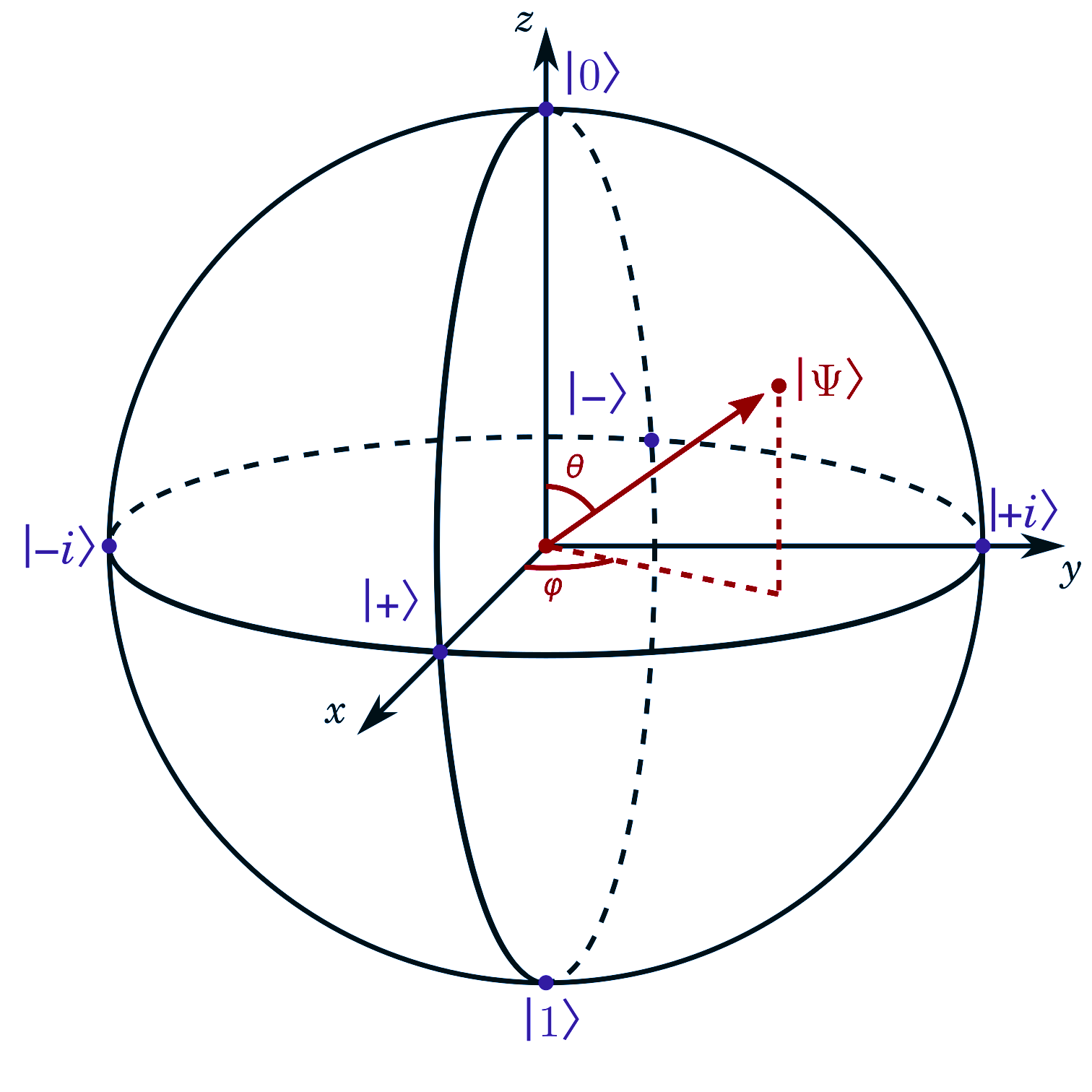}
    \caption{A Bloch sphere representation of a single qubit. The north and south poles typically correspond to the computational basis states $\ket{0}$ and $\ket{1}$, respectively. Any pure qubit state can be visualized as a point on the sphere, with the polar angle $\theta$ and azimuthal angle $\phi$ specifying its amplitude and phase. This geometric picture highlights how superposition and relative phase are captured by positions on the sphere’s surface.}
\label{fig:bloch}
\end{figure*}

\subsection{Decoherence and Noise}
\label{subsec:decoherence-background}

Decoherence is the process by which a qubit loses its quantum character, superposition and entanglement, through unwanted interaction with its environment, causing a pure state to evolve into a mixed state \cite{NielsenChuang,TurroThesis}. In superconducting qubits specifically, decoherence arises from several coupled mechanisms:

\begin{itemize}
    \item \textbf{Quasiparticle tunneling}, in which thermally or radiatively generated excitations in the superconductor tunnel across the junction and disturb the qubit's energy levels \cite{Catelani2012,MartinisOsborne}.
    \item \textbf{Charge and flux noise}, arising from fluctuating electric charge or magnetic flux in the qubit's local environment \cite{MartinisOsborne,TurroThesis}.
    \item \textbf{Photon loss and spontaneous emission}, relevant to any platform that stores or transmits information optically, and to resonator-coupled superconducting qubits specifically
    \cite{NielsenChuang}.
\end{itemize}

Decoherence is quantified by two characteristic timescales: $T_1$, the energy relaxation time, and $T_2$, the phase coherence time. Both must be long compared to the time required to execute a gate for computation to be practical. Rather than duplicate the discussion here, we defer strategies for extending $T_1$ and $T_2$, qubit design, materials choices, and operational techniques, to Section~\ref{sec:challenges}, where they are addressed alongside the scalability and error-correction challenges they are meant to solve.

\subsection{The Qubit: Representation and Purity}

A single qubit's state can be visualized as a point on the \emph{Bloch sphere} (Figure~\ref{fig:bloch}), parameterized by two angles:
\[
|\psi\rangle = \cos\!\left(\frac{\theta}{2}\right)|0\rangle +
e^{i\phi}\sin\!\left(\frac{\theta}{2}\right)|1\rangle,
\]
where $\theta$ is the polar angle from $|0\rangle$ and $\phi$ is the azimuthal phase \cite{NielsenChuang, BarencoEtAl,VatanWilliams}. Every \emph{pure} state, one fully described by a single state vector, sits on the sphere's surface. A \emph{mixed} state, representing classical uncertainty about which pure state a system occupies (for example, from
unwanted environmental entanglement), corresponds to a point in the sphere's interior and requires a density matrix,
\[
\rho = \sum_i p_i |\psi_i\rangle\langle\psi_i|,
\]
for its description, where $p_i$ is the classical probability of occupying pure state $|\psi_i\rangle$ \cite{TurroThesis}. The purity of a state is quantified by $\mathrm{Tr}(\rho^2)$, equal to 1 for a pure state and less than 1 for a mixed state. This formalism reappears in Section~\ref{sec:challenges}, where decoherence is described as the drift of a qubit's state from the surface of the Bloch sphere toward its center, and where a logical qubit's density matrix must be tracked across many physical qubits simultaneously.

With these foundations in place, we now turn to the physical mechanisms, quantum tunneling, Cooper pair formation, and the Josephson effect, that allow a real superconducting circuit to realize a controllable qubit.

\section{Fundamentals of Constructing a QPU}
\label{sec:fundamentals}

\begin{figure*}[t]
    \centering
    \includegraphics[width=0.6\textwidth]{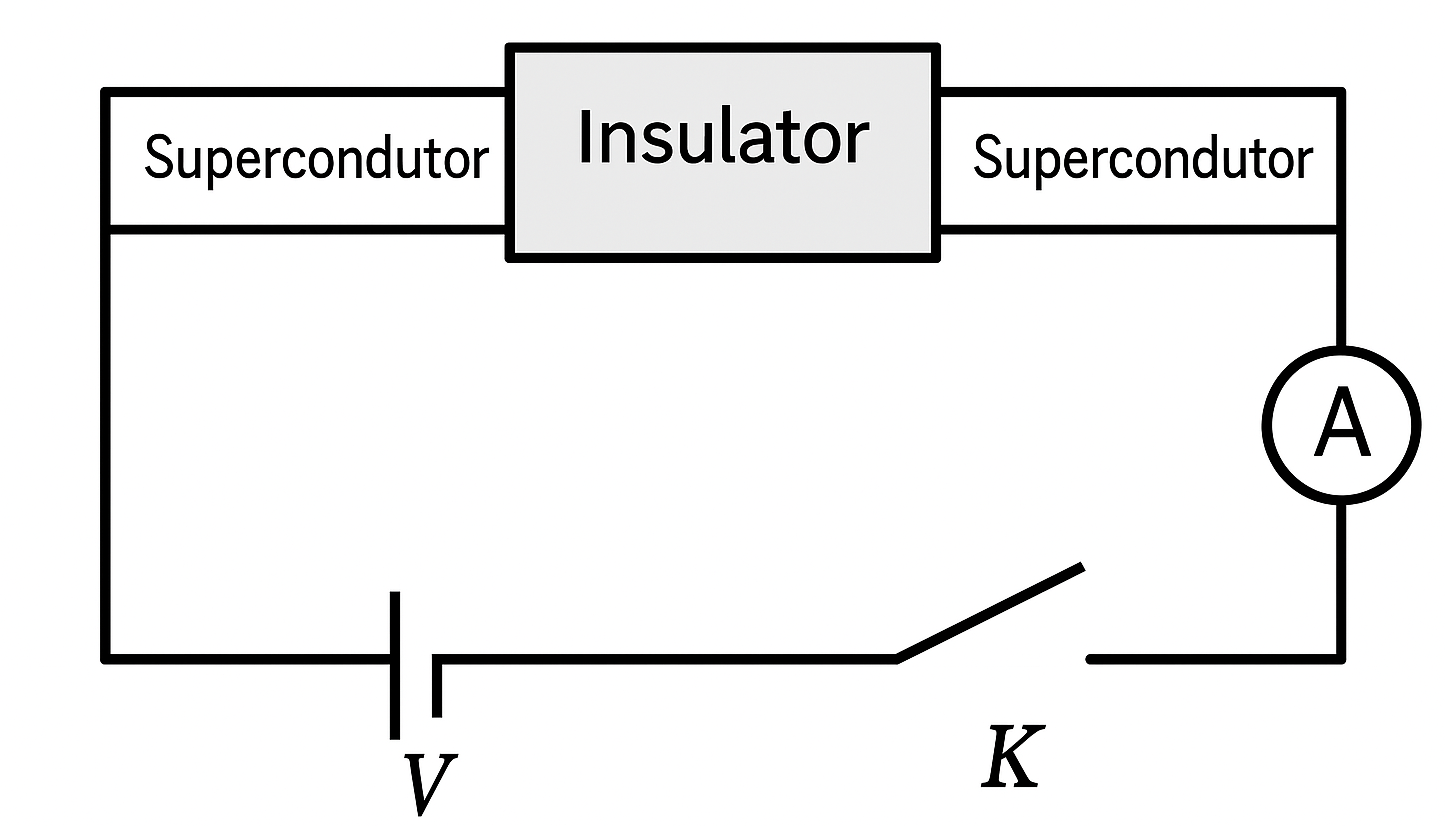}
    \caption{This diagram illustrates a Josephson junction, which consists of two superconductors \(S_1\) and \(S_2\) separated by a thin insulating barrier \(Insulator\). The circuit includes a voltage source \(V\) to apply a potential difference, a switch \(K\) to control the current flow, and an ammeter \(A\) to measure the resulting current. }
\label{fig:jj-circuit}
\end{figure*}

Building a QPU means engineering a physical system whose quantum properties can be initialized, controlled, and measured on demand. The dominant approach uses superconducting circuits built around \emph{Josephson junctions}: two superconductors separated by a thin insulating barrier (Figure~\ref{fig:jj-circuit}). At millikelvin temperatures, electron pairs tunnel coherently across this barrier, producing quantum behavior that can be shaped into a controllable qubit \cite{ClarkeBraginski}. This section develops that behavior from the ground up: quantum tunneling, Cooper pair formation, the Josephson effects, and finally the circuit design that turns a junction into a usable qubit.

\subsection{Quantum Tunneling}

Quantum tunneling is the phenomenon by which a particle has a nonzero probability of appearing on the far side of an energy barrier it does not classically have enough energy to cross. This follows from the wave-like nature of matter: a particle's wavefunction does not drop to zero inside a classically forbidden region, only decays, so it retains finite amplitude on the other side \cite{NielsenChuang}. The tunneling probability falls off as the barrier grows thicker or higher, but never reaches exactly zero. This is the mechanism that allows Cooper pairs to cross the insulating layer of a Josephson junction without dissipating energy, and it is the physical basis for everything that follows in this section.

\subsection{Cooper Pair Formation}

\begin{figure*}[t]
    \centering
    \includegraphics[width=0.5\textwidth]{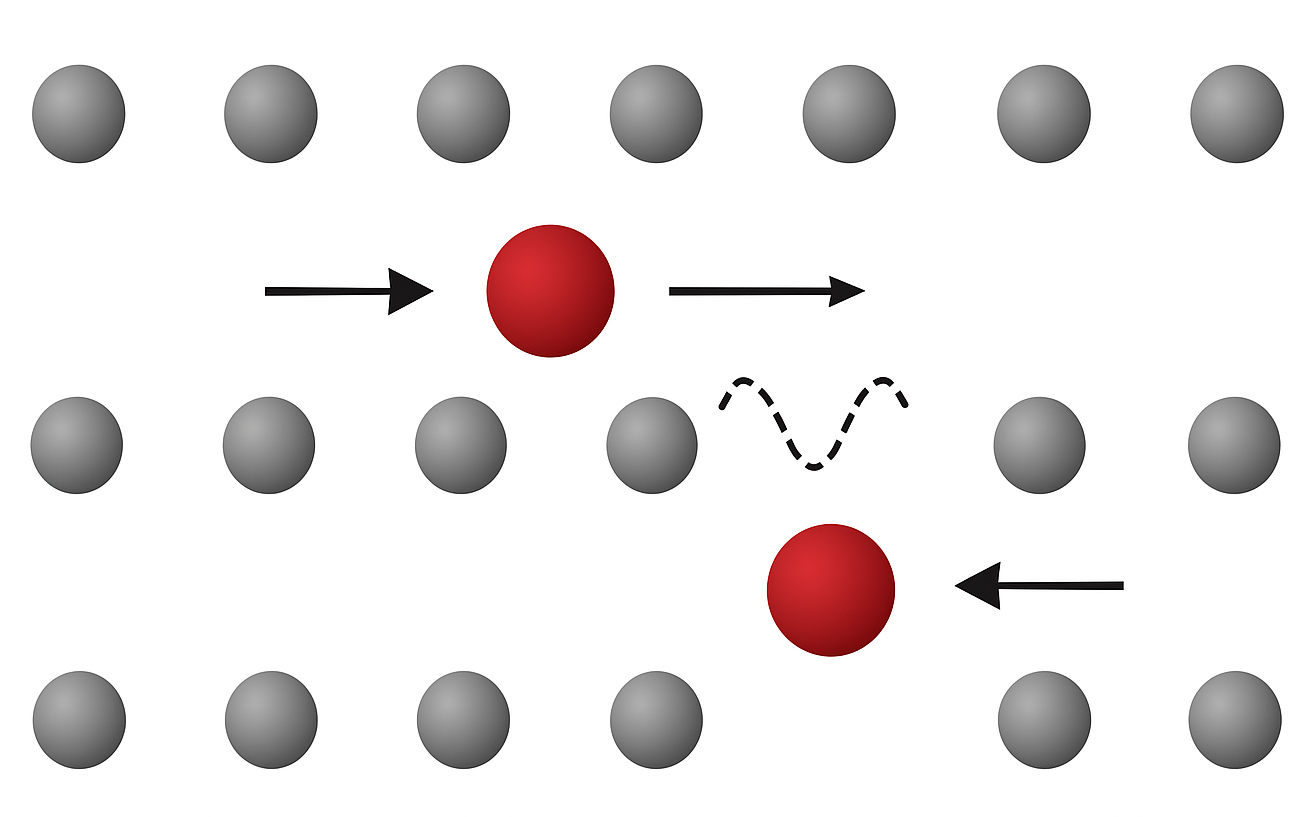}
    \caption{A simplified schematic of Cooper pair formation in a superconductor. In this illustration, the gray spheres represent lattice ions, forming the stationary background of the material. The red colored spheres depict the electrons. These electrons carry opposite spins and momenta. The wavy line between them symbolizes phonon vibrations—dynamic lattice interactions that mediate the attractive force binding the electrons together into a Cooper pair. This pairing enables the electrons to travel through the lattice without resistance.}
\label{fig:cooper-pair}
\end{figure*}

A Cooper pair (Figure~\ref{fig:cooper-pair}) is a bound state of two electrons in a superconductor, mediated by an attractive interaction with the crystal lattice. In an ordinary conductor, electrons repel one another electrostatically; in a superconductor, an electron moving through the lattice induces a slight, transient distortion that attracts a second electron, coupling the pair via lattice vibrations (phonons) \cite{Tinkham}. This phonon-mediated attraction is strong enough to overcome Coulomb repulsion at sufficiently low temperature. Cooper pairs form from electrons of opposite momentum and opposite spin, giving the pair a net spin of zero and allowing it to behave as a boson rather than a fermion \cite{Tinkham}, a property essential to BCS theory and to the collective, dissipationless flow of current in a superconductor.

\subsection{The Josephson Effects}

Named for Brian Josephson, the Josephson effects describe how a supercurrent behaves across a thin insulating barrier separating two superconductors. Writing the superconducting wavefunctions on either side of the junction as $\psi_1 = n_1 e^{i\theta_1}$ and $\psi_2 = n_2
e^{i\theta_2}$, with phase difference $\delta = \theta_1 - \theta_2$ \cite{Suzuki2012}, the continuity equation for Cooper-pair density gives:
\[
\frac{\partial n_1}{\partial t} = -2Tn_2\sin(\delta), \qquad
\frac{\partial n_2}{\partial t} = 2Tn_1\sin(\delta),
\]
where $T$ is the interaction strength coupling the two superconductors. Taking $n_1 \approx n_2 = n$ and noting that current is proportional to $\partial n/\partial t$ yields the \textbf{DC Josephson effect}:
\begin{equation}
I_s = I_c \sin(\phi)
\label{eq:dc-josephson}
\end{equation}
where $I_c$ is the critical current, the maximum supercurrent the junction can sustain without resistance, and $\phi$ is the superconducting phase difference across the junction. Equation \eqref{eq:dc-josephson} shows that a steady supercurrent flows across the junction with \emph{no applied voltage}, driven entirely by quantum tunneling of Cooper pairs and set by the phase difference $\phi$ \cite{ClarkeBraginski}.

When a voltage $V$ is applied across the junction, the phase evolves in time according to
\begin{equation}
\frac{\partial \phi}{\partial t} = \frac{2eV}{\hbar}, \qquad \hbar = \frac{h}{2\pi},
\label{eq:phase-evolution}
\end{equation}
a relation that follows from applying the time-dependent Schr\"odinger equation to the phase of the junction wavefunction. Since the oscillation frequency is $f = \frac{1}{2\pi}\frac{\partial \phi}{\partial t}$, substituting Equation~\eqref{eq:phase-evolution} gives the \textbf{AC Josephson effect}:
\begin{equation}
f = \frac{2eV}{h}.
\label{eq:ac-josephson}
\end{equation}
The supercurrent therefore oscillates at a frequency directly proportional to the applied voltage, the basis of both Josephson voltage standards (Section~\ref{sec:applications}) and the microwave-frequency control used to drive superconducting qubits.

\paragraph{Example.} For $V = 1\,\mu\text{V}$,
\[
f = \frac{2 \times (1.602\times10^{-19}\,\text{C}) \times (1\times10^{-6}\,\text{V})}{6.626\times10^{-34}\,\text{J s}}
\]
\[
\approx 4.83\times10^{8}\,\text{Hz} = 4.83\,\text{GHz},
\]
a frequency typical of the microwave drives used to operate superconducting qubits in practice.

The phase difference $\phi$ is constrained to be single-valued modulo $2\pi$ at every point in the superconductor \cite{Suzuki2012}; in circuits with multiple junctions, the total phase difference around a loop is the sum of the phase differences across each junction, a fact used directly in flux-based qubit designs (Section~\ref{sec:challenges}). In the DC regime ($V=0$), $\phi$ is static and the supercurrent is steady; in the AC regime, $\phi(t) = \phi_0 + \tfrac{2eV}{\hbar}t$ evolves linearly with time, producing an oscillatory current.

\subsection{From Junction to Qubit: Circuit Realizations}
\label{subsec:junction-to-qubit}

A single junction on its own is not yet a qubit, it is a nonlinear inductor. To obtain a controllable two-level system, the junction is shunted by a capacitor, forming a nonlinear $LC$ oscillator whose energy levels are \emph{unevenly spaced}, unlike the evenly spaced levels of a simple harmonic oscillator. This anharmonicity is what allows the lowest two levels, $|0\rangle$ and $|1\rangle$, to be addressed selectively with a microwave drive without unintentionally exciting higher levels \cite{Krantz2019}. The circuit's Hamiltonian takes the approximate form
\begin{equation}
\hat{H} \approx 4E_C\hat{n}^2 - E_J\cos\hat{\phi},
\label{eq:transmon-hamiltonian}
\end{equation}
where $E_C$ is the charging energy set by the shunt capacitance, $E_J$ is the Josephson energy set by the junction's critical current, $\hat{n}$ is the Cooper-pair number operator, and $\hat{\phi}$ is the phase operator conjugate to $\hat{n}$. The ratio $E_J/E_C$ is the single most consequential design choice in a superconducting qubit:

\begin{itemize}
    \item Operating with $E_J/E_C \gg 1$ (the \textbf{transmon} regime) exponentially suppresses the qubit's sensitivity to charge noise at the cost of reduced anharmonicity, and is the design used in the majority of superconducting QPUs deployed today \cite{Krantz2019}. The motivation for this design choice, and its trade-offs, are discussed further in Section~\ref{sec:challenges}.
    \item An alternative, the \textbf{fluxonium}, shunts the junction with a large additional inductance rather than relying solely on charging energy. This suppresses charge and flux sensitivity simultaneously, yielding coherence times reported to exceed those of comparable transmons in dedicated experiments, at the cost of a more complex fabrication and control scheme \cite{manucharyan2009,nguyen2019}. Fluxonium has grown from a research curiosity into a serious contender for reducing the physical-to-logical qubit ratio in error-corrected architectures.
\end{itemize}

Once realized as an anharmonic circuit, the qubit is manipulated with calibrated microwave pulses. Single-qubit gates, the Pauli-$X$, Pauli-$Y$, and Hadamard gates among them, are implemented by applying pulses resonant with the $|0\rangle \leftrightarrow |1\rangle$ transition; pulse duration sets the rotation angle on the Bloch sphere.

\begin{figure*}[t]
    \centering
    \includegraphics[width=0.7\textwidth]{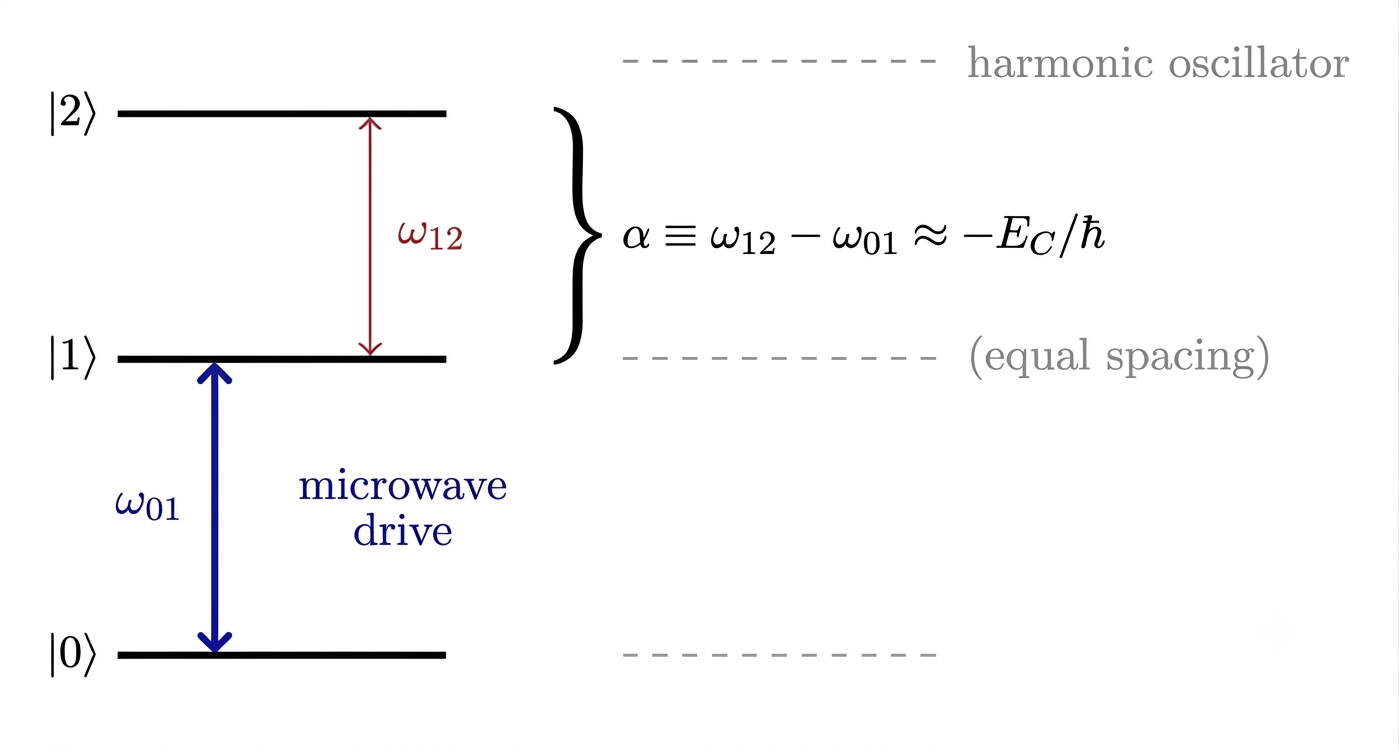}
    \caption{Energy-level structure of a transmon qubit (solid) compared to an ideal harmonic oscillator (dashed). Anharmonicity $\alpha = \omega_{12}-\omega_{01}$ detunes the $|1\rangle\to|2\rangle$ transition from the microwave drive resonant with $|0\rangle\leftrightarrow|1\rangle$, allowing the qubit subspace to be addressed selectively.}
    \label{fig:energy-levels}
\end{figure*}

\paragraph{State Preparation via Resonant Driving.}
The connection between the abstract superposition of Section~\ref{sec:background} and the physical circuit developed above is made explicit by how a microwave pulse actually acts on the qubit. In the two-level truncation of Equation~\eqref{eq:transmon-hamiltonian}, the bare qubit Hamiltonian is $\hat{H}_0 = \tfrac{\hbar\omega_{01}}{2}\hat{\sigma}_z$, where the transition frequency $\omega_{01}$ between $|0\rangle$ and $|1\rangle$ is set directly by the circuit parameters,
\begin{equation}
\omega_{01} \approx \frac{\sqrt{8E_JE_C}-E_C}{\hbar},
\label{eq:qubit-frequency}
\end{equation}
in the transmon limit $E_J/E_C \gg 1$ \cite{koch2007}. A classical microwave drive capacitively coupled to the qubit adds a term $\hat{H}_d(t) = \hbar\Omega(t)\cos(\omega_d t+\varphi)\hat{\sigma}_x$ to the Hamiltonian, where $\Omega(t)$ is set by the drive amplitude. When the drive is tuned to resonance, $\omega_d = \omega_{01}$, this is precisely the condition needed to selectively address the $|0\rangle\leftrightarrow |1\rangle$ transition without exciting $|2\rangle$ (Figure~\ref{fig:energy-levels}), the anharmonicity established in Equation~\eqref{eq:transmon-hamiltonian} is what keeps $\omega_{12}$ detuned from $\omega_{01}$ far enough for this selectivity to hold.

Under resonant driving, moving to a frame rotating at $\omega_d$ and applying the rotating-wave approximation gives Rabi oscillation: a qubit initialized in $|0\rangle$ evolves as
\begin{equation}
\begin{split}
|\psi(t)\rangle &= \cos\!\left(\frac{\theta(t)}{2}\right)|0\rangle
- i\sin\!\left(\frac{\theta(t)}{2}\right)e^{-i\varphi}|1\rangle, \\
\theta(t) &= \int_0^t \Omega(t')\,dt',
\end{split}
\label{eq:rabi}
\end{equation}
where $\theta(t)$ is the pulse area and $\Omega$ is the Rabi frequency. Equation~\eqref{eq:rabi} is the physical realization of the amplitudes $\alpha,\beta$ introduced in Section~\ref{sec:background}: the pulse duration and phase directly set $\alpha=\cos(\theta/2)$ and $\beta=-i\sin(\theta/2)e^{-i\varphi}$. A \emph{$\pi$-pulse}, with $\theta=\pi$, fully inverts the qubit population ($|0\rangle\to |1\rangle$); a \emph{$\pi/2$-pulse}, with $\theta=\pi/2$, stops halfway and leaves the qubit in the equal superposition $\tfrac{1}{\sqrt2}(|0\rangle - i|1\rangle)$. This is the concrete mechanism by which a Josephson-junction qubit is placed into superposition on demand, and it is the operation underlying every single-qubit gate discussed below.

\paragraph{Example.} A typical Rabi frequency for a capacitively driven transmon is $\Omega/2\pi \approx 20\,\text{MHz}$ \cite{Krantz2019}. Preparing a $\pi/2$-pulse then requires a pulse duration
\[
t_{\pi/2} = \frac{\pi/2}{\Omega} = \frac{1}{8\times(20\times10^6\,\text{Hz})} \approx 12.5\,\text{ns},
\]
consistent with the tens-of-nanoseconds single-qubit gate times typical of current superconducting hardware.

Before any drive is applied, the qubit must first be prepared in a known state. In practice this relies on the qubit's transition energy being large compared to the thermal energy at millikelvin operating temperatures, $\hbar\omega_{01} \gg k_BT$, so that passive thermalization leaves the qubit in $|0\rangle$ with high probability; active reset protocols using engineered dissipation are also used where faster or
higher-fidelity initialization is required \cite{Krantz2019}.

Two-qubit gates require coupling between qubits, typically mediated by a shared microwave resonator or a tunable coupler, and are used to generate the entanglement that underlies quantum algorithms \cite{Krantz2019}. Precise pulse shaping is essential: pulses that are too abrupt spectrally leak into neighboring transitions, an effect that must be actively suppressed given the qubit's finite anharmonicity. This tension between control precision and physical device design reappears throughout the challenges discussed next.

\paragraph{Readout: Distinguishing $|0\rangle$ from $|1\rangle$.} Measurement is performed indirectly, by coupling each qubit to its own linear readout resonator (a coplanar-waveguide cavity) rather than measuring the qubit directly. The qubit-resonator coupling strength $g$ and detuning $\Delta = \omega_r - \omega_{01}$ are chosen so that $|\Delta| \gg g$, the \emph{dispersive regime}, in which the qubit and resonator do not exchange energy but the resonator's frequency instead acquires a qubit-state-dependent shift,
\begin{equation}
\omega_r \;\longrightarrow\; \omega_r \pm \chi, \qquad
\chi \approx \frac{g^2}{\Delta}\left(\frac{\alpha}{\Delta-\alpha}\right),
\label{eq:dispersive-shift}
\end{equation}
with the sign of $\chi$ set by whether the qubit is in $|0\rangle$ or $|1\rangle$, and $\alpha$ the anharmonicity from
Section~\ref{subsec:junction-to-qubit} \cite{koch2007}. Crucially, this shift exists \emph{because} the transmon is anharmonic, a perfectly harmonic qubit would produce no state-dependent shift at all, so the same anharmonicity that enables selective driving (Figure~\ref{fig:energy-levels}) is also what makes readout possible.

\begin{figure*}[h]
    \centering
    \includegraphics[width=0.5\textwidth]{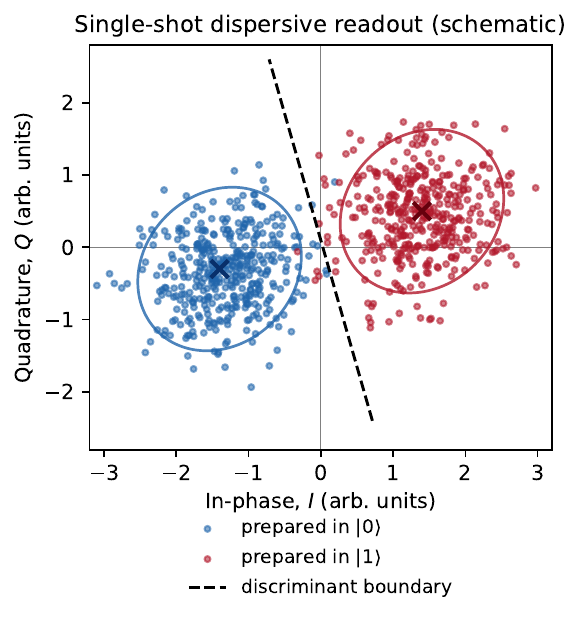}
    \caption{Schematic single-shot readout histogram. Repeated
    measurements of a qubit prepared in $|0\rangle$ (blue) or $|1\rangle$ (red) cluster at different points in the IQ plane because of the state-dependent dispersive shift $\pm\chi$; a discriminant boundary (dashed) classifies each individual shot.}
    \label{fig:readout}
\end{figure*}

To read the qubit out, a weak microwave tone is sent toward the resonator at a fixed probe frequency near $\omega_r$, and the amplitude and phase of the reflected or transmitted signal are recorded with a heterodyne/homodyne receiver. Because the resonator's true resonance sits at $\omega_r+\chi$ or $\omega_r-\chi$ depending on the qubit state, the returned signal picks up a different phase and amplitude in each case. Repeating this for many identical preparations and plotting each outcome as a point in the complex (in-phase, quadrature) plane produces two separated clusters, one per qubit state (Figure~\ref{fig:readout}); a simple threshold or, more precisely, a linear discriminant between the two clusters converts each single measurement into a classified bit. Well optimized dispersive readout on current hardware routinely achieves single-shot fidelities above 99\% within a few hundred nanoseconds to roughly a microsecond of integration time \cite{Krantz2019}. This is also where the readout-line and multiplexing infrastructure discussed in Sections~\ref{subsec:control-complexity} and~\ref{sec:challenges} connects back to the physical qubit layer: each resonator's transmission line is exactly what must be wired, filtered, and ultimately multiplexed as qubit count grows.

\section{Challenges in Scaling Josephson-Junction QPUs}
\label{sec:challenges}

Having established how a single Josephson-junction qubit is built and controlled, we now turn to what happens when hundreds, thousands, or millions of them must work together. Three problems compound as QPUs grow: unwanted interactions between neighboring qubits, the sheer engineering burden of controlling each qubit individually from room temperature, and the fact that decoherence and gate errors do not simply average out but must be actively corrected. This section addresses each in turn, closing with the error-correction overhead required to turn noisy physical qubits into a smaller number of reliable logical ones.

\subsection{Crosstalk}

As qubit count increases, qubits are typically packed closer together on chip, and the physical distance available to isolate one qubit's control and readout lines from its neighbors' shrinks accordingly. Unwanted microwave coupling between nearby qubits, crosstalk, degrades gate fidelity and complicates the assumption that single-qubit operations can be run in parallel across a chip \cite{Rebello2024}. Multiplexing readout, where a single feedline services several resonators each coupled to a different qubit, reduces the physical cabling burden but introduces its own crosstalk-management problem, since signals meant for one qubit can leak into the readout channel of another \cite{Rebello2024}. Mitigating crosstalk increasingly drives chip layout itself: qubit frequencies must be allocated across the chip so that neighboring qubits are detuned enough to avoid unwanted resonant coupling, a constraint that grows harder to satisfy as qubit count increases.

\subsection{Control Complexity and the Classical Quantum Interface}
\label{subsec:control-complexity}

Every physical qubit needs a microwave control line, and in most current architectures, its own readout line as well. At the scale of tens or a few hundred qubits, this is manageable with individual coaxial cables running from room-temperature electronics down into the dilution refrigerator. At the scale needed for fault-tolerant computation, plausibly hundreds of thousands to millions of physical qubits, once error-correction overhead is included (Section~\ref{subsec:qec}), this wiring approach does not scale. Each additional cable adds physical footprint inside a fridge with famously limited space, adds heat load that must be removed at progressively colder and more expensive cooling stages, and adds cost and calibration burden \cite{Krantz2019}.

Two complementary strategies are being pursued to close this gap:

\begin{itemize}
    \item \textbf{Cryo-CMOS control electronics} move part of the control and readout circuitry from room temperature into the cryostat itself, operating at intermediate temperature stages (around 3-4~K) where conventional CMOS transistors still function correctly. This drastically reduces the number of cables that must cross the largest temperature gradient, at the cost of designing electronics that tolerate cryogenic operation and the reduced power budget available at those stages \cite{patra2018}.
    \item \textbf{Multiplexed and photonic I/O} pushes further by combining many control or readout signals onto a single physical line, whether through frequency-domain multiplexing of microwave signals or, more recently, by converting control and readout signals to optical photons that can be carried on a single fiber with far lower heat load than an equivalent bundle of coaxial cables \cite{Krantz2019}.
\end{itemize}

Beyond wiring, the classical compilation and control-logic burden itself grows with qubit count: circuit compilation must decompose increasingly large algorithms into physically realizable single and two-qubit gate sequences, and the classical control system must orchestrate the timing of potentially millions of interacting physical qubits with the precision required for error correction. Managing the interface between the physical qubit layer and the classical apparatus that drives it is, in practice, as much a systems-engineering problem as it is a physics problem \cite{NielsenChuang}, and is the reason large-scale roadmaps now report progress in classical I/O density alongside qubit count and gate fidelity.

\subsection{Decoherence and Noise Mitigation at Scale}
\label{subsec:decoherence-mitigation}

Section~\ref{subsec:decoherence-background} introduced decoherence as the loss of a qubit's quantum information through environmental interaction, and Section~\ref{subsec:junction-to-qubit} noted that operating in the transmon regime ($E_J/E_C \gg 1$) exponentially suppresses sensitivity to charge noise. That single design choice is the starting point for coherence engineering, not the end of it. In practice, several further strategies are combined:

\begin{itemize}
    \item \textbf{Balancing anharmonicity against noise protection.} Increasing $E_J/E_C$ further reduces charge-noise sensitivity but linearly reduces anharmonicity, making it easier to accidentally drive transitions to higher energy levels; qubit design is a continual trade-off between these two effects \cite{Krantz2019}.
    \item \textbf{Filtering and shielding.} Infrared radiation reaching the qubit chip can generate quasiparticles and accelerate their tunneling across junctions, a dominant decoherence channel \cite{Catelani2012}; RF and IR filtering, electromagnetic shielding, and careful thermal anchoring inside the dilution refrigerator all suppress this \cite{Rebello2024}.
    \item \textbf{Calibration.} Microwave mixers used to generate qubit control pulses must be calibrated against amplitude and phase imbalance and local-oscillator leakage, both of which introduce gate errors if left uncorrected \cite{jolin2020}.
    \item \textbf{Optimal control.} Pulse shapes are optimized, rather than left as simple square or Gaussian envelopes, to minimize leakage into higher energy levels given a qubit's finite anharmonicity \cite{TurroThesis}.
    \item \textbf{Fabrication reproducibility.} Junction-to-junction variation during fabrication is itself a source of effective noise at the chip level, since it produces qubit frequencies that deviate from design targets. Improving the reproducibility of electron-beam lithography, including dose control and proximity-effect correction, directly improves yield and coherence consistency across a chip \cite{Rebello2024}.
    \item \textbf{Materials choice.} Aluminum-based junctions have historically shown longer coherence than niobium-based junctions, plausibly linked to how each material's superconducting gap responds to oxygen incorporation and scattering defects \cite{MartinisOsborne}; Section~\ref{sec:applications} discusses renewed interest in
    niobium as fabrication techniques have improved.
    \item \textbf{Flux qubits specifically} require careful tuning of loop inductance relative to junction energy to achieve the intended nonlinearity, and often use deliberately small-area junctions to shape their noise sensitivity \cite{MartinisOsborne}.
\end{itemize}

These measures extend coherence times, but they do not eliminate errors, they buy the additional error budget that error correction, discussed next, needs in order to work at all.

\subsection{Quantum Error Correction Overhead}
\label{subsec:qec}

No amount of materials engineering or pulse shaping removes errors entirely; a fault-tolerant quantum computer instead needs an error-correction scheme that can detect and correct errors faster than they accumulate. This requires encoding one \emph{logical} qubit's worth of protected information across several \emph{physical} qubits.

\subsubsection{Small Codes and the Concatenation Idea}

The simplest error-correcting codes illustrate the underlying idea without being the ones actually deployed in current hardware. The Steane code encodes one logical qubit into seven physical qubits; the 9-qubit Shor code does the same with nine \cite{NielsenChuang}. To correct $t$ errors, the code's distance, the minimum number of physical errors needed to cause an undetected logical error, must satisfy $d \geq 2t+1$. Concatenating a code with itself, encoding logical qubits of logical qubits, increases distance (and hence tolerable error rate) exponentially at the cost of an exponential increase in physical qubit count, $n^L$ for $L$ levels of concatenation using an $n$-qubit code \cite{NielsenChuang}. This is the origin of the Threshold Theorem: if the physical error rate is below a critical threshold, arbitrarily reliable logical computation becomes possible by adding enough physical qubits \cite{NielsenChuang}.

\subsubsection{The Surface Code}

\begin{figure*}[t!]
    \centering
    \includegraphics[width=0.55\textwidth]{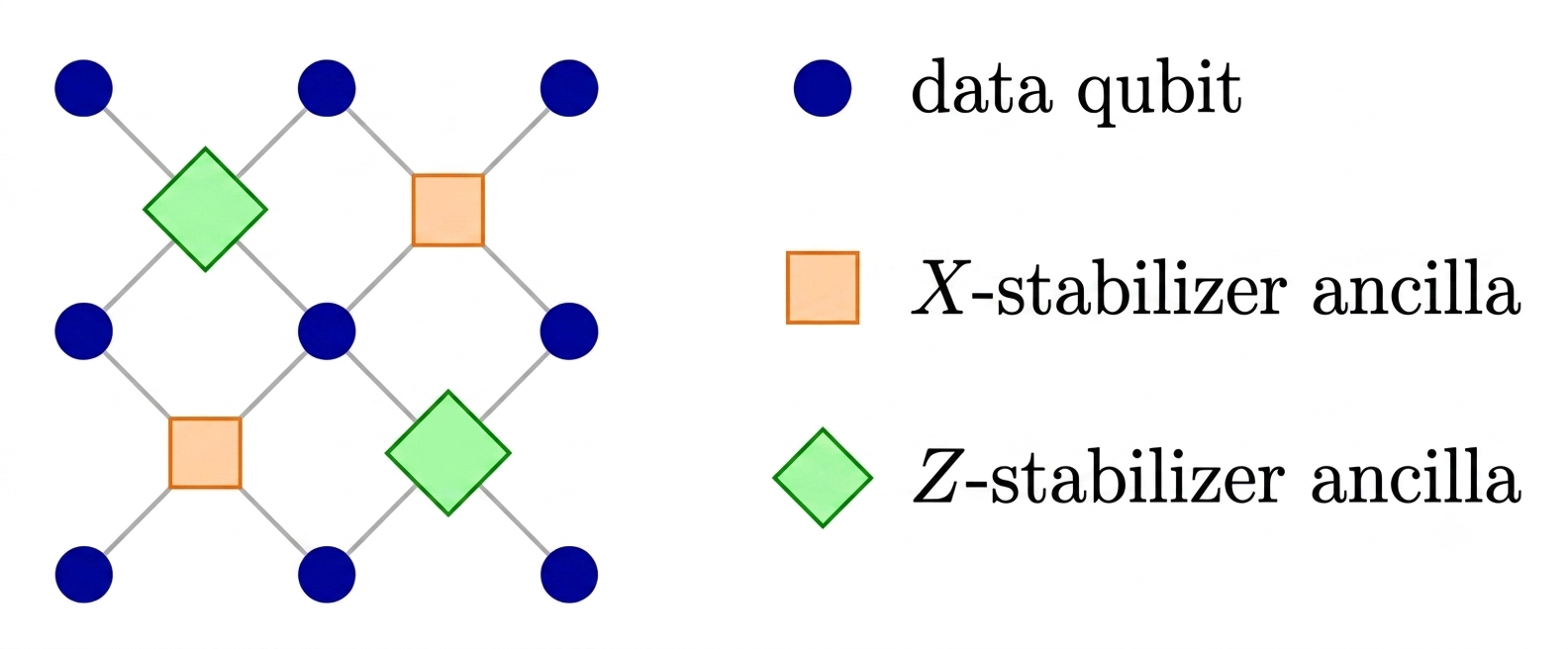}
    \caption{Simplified schematic of a distance-3 surface code patch. Data qubits (blue) sit on a square lattice; ancilla qubits measure either $X$-type (orange) or $Z$-type (green) stabilizers on alternating plaquettes, each acting on its four neighboring data qubits. Boundary/weight-2 ancillas are omitted for clarity.}
    \label{fig:surface-code}
\end{figure*}

Superconducting hardware does not use Steane-type codes in practice. Its native qubit connectivity, each qubit coupled only to its nearest neighbors on a two-dimensional chip, matches the \emph{surface code} far better: a topological code defined on a 2D lattice of physical qubits, where errors are detected by repeatedly measuring local stabilizer operators without directly measuring (and thus destroying) the encoded logical information (Figure~\ref{fig:surface-code}) \cite{fowler2012}. Surface-code distance scales with the linear size of the physical qubit patch used to encode one logical qubit, and unlike Steane-type codes, it tolerates a comparatively high physical error threshold, making it the practical choice for near-term superconducting hardware.

This is no longer a purely theoretical proposal. In 2024-2025, Google Quantum AI demonstrated two surface-code memories on their Willow processor operating \emph{below} the surface-code threshold for the first time: a distance-5 code with a real-time decoder, and a 101-qubit distance-7 code achieving a logical error rate of 0.143\% per correction cycle, with each increase of two in code distance suppressing the logical error rate by a factor of roughly 2.14 \cite{acharya2025willow}. That distance-7 logical qubit exceeded the lifetime of its single best physical qubit by a factor of about 2.4, the first experimental demonstration of a logical qubit meaningfully outperforming any of its physical constituents \cite{acharya2025willow}. The result confirms the threshold theorem experimentally for superconducting hardware, but it also makes the remaining gap concrete: even at this result's measured error rate, several more orders of magnitude of suppression are still needed before error-corrected superconducting hardware can run algorithms of practical size \cite{acharya2025willow}.

\subsubsection{Overhead and the Push Toward Lower-Overhead Codes}

The central cost of the surface code is physical-qubit overhead: current demonstrations require on the order of 100 physical qubits per logical qubit at the code distances needed for useful error suppression, and fault-tolerant algorithms are expected to need hundreds to thousands of logical qubits, implying physical qubit counts in the hundreds of thousands to millions once ancilla qubits for syndrome extraction are included \cite{NielsenChuang,acharya2025willow}. This overhead has motivated active research into \emph{quantum low-density parity-check (LDPC) codes}, which use longer-range qubit connectivity to achieve comparable error suppression with substantially fewer physical qubits per logical qubit than the surface code, at the cost of requiring non-local couplings that are harder to realize in a purely two-dimensional superconducting layout \cite{bravyi2024}. Whether qLDPC codes or the surface code (or some hybrid) ultimately dominates large-scale superconducting architectures remains an open systems-design question, trading connectivity complexity against qubit-count overhead.

\subsubsection{System-Level Consequences}

Whichever code is used, encoding logical qubits reshapes the entire QPU architecture, not just the qubit layer:

\begin{itemize}
    \item \textbf{Ancilla management.} Error correction requires ancilla qubits to extract error syndromes without collapsing the encoded data; preparing, verifying, and re-using these ancillas efficiently is itself a significant design problem, addressed in part by the control-interface strategies discussed in Section~\ref{subsec:control-complexity} \cite{NielsenChuang}.
    \item \textbf{Architectural specialization.} Large error-corrected designs may use different codes or qubit densities for computation versus memory regions, requiring transfer networks that move (and potentially re-encode) quantum data between regions without decoding it \cite{NielsenChuang}.
    \item \textbf{Long-range communication.} Moving logical information across a large device, or between separate QPU modules, is increasingly proposed to rely on quantum teleportation over purified entangled pairs, reducing the error propagation risk of direct physical qubit transport \cite{NielsenChuang}.
    \item \textbf{Design automation.} Scaling circuit design to error-corrected devices with these architectural constraints is driving development of dedicated electronic design automation (EDA) tooling for quantum circuits, analogous to classical chip design flows \cite{Ayala2023}.
    \item \textbf{Flux trapping.} At the physical layer, unwanted magnetic flux trapped in superconducting films degrades performance; moat structures that deliberately attract stray flux away from sensitive circuit regions are one mitigation used in current chip layouts \cite{Ayala2023}.
\end{itemize}

Balancing the resources spent on error correction against the resources available for computation, much of the chip is qubits doing useful work versus qubits protecting that work, is ultimately an application-driven decision: the level of encoding needed depends on algorithm size and the reliability it demands \cite{NielsenChuang}. The Willow result marks the point at which this balance became empirically measurable for superconducting hardware rather than purely theoretical; the next section turns to where that leaves near-term applications and ongoing materials research.

\section{Applications and Future Perspectives}
\label{sec:applications}

Josephson junctions have moved well beyond quantum computing as their sole application; they underpin some of the most sensitive measurement instruments available today, and ongoing materials-science work continues to reshape which qubit designs are viable at scale. This section surveys established applications, summarizes recent results demonstrating near-term computational value, reviews the materials advances driving coherence improvements, and closes with hybrid and emerging directions.

\subsection{Established Applications}

\subsubsection{SQUIDs}

A Superconducting Quantum Interference Device (SQUID) uses two Josephson junctions embedded in a superconducting loop to detect magnetic flux with extraordinary sensitivity. The Josephson effect establishes a periodic relationship between the current through the loop and the magnetic flux threading it, allowing SQUIDs to resolve flux changes far smaller than those detectable by any classical magnetometer \cite{ClarkeBraginski}. This sensitivity underlies applications from biomagnetic imaging to geophysical surveying, well outside the scope of quantum computing itself.

\subsubsection{Voltage Standards}

Arrays of Josephson junctions, driven by microwave radiation at a well-defined frequency, produce precisely quantized voltage steps tied directly to fundamental constants via Equation~\eqref{eq:ac-josephson}. This makes Josephson junction arrays the metrological basis for calibrating voltage-measuring instruments with a precision unattainable through classical means \cite{ClarkeBraginski}, and they remain the international standard for voltage realization.

\subsubsection{Quantum Computing}

As developed in Sections~\ref{sec:fundamentals} and~\ref{sec:challenges}, Josephson junctions provide the nonlinear circuit element needed to realize an addressable qubit: they enable representation of quantum information, single- and two-qubit gate operations via microwave and flux control, and dispersive readout of qubit state through coupled resonators \cite{Shor,CaoRomeroAspuru,ClarkeBraginski}. This remains the highest-profile application of the technology and the focus of the remainder of this paper.

\subsection{Near-Term Computational Results}
\label{subsec:near-term}

Superconducting QPUs have begun producing results that are difficult to reproduce with brute-force classical simulation, even though fault-tolerant operation remains out of reach. In 2023, IBM reported measurements of expectation values for a kicked Ising model on a 127-qubit superconducting processor at circuit depths that, at the time, exceeded the reach of leading classical tensor-network methods \cite{kim2023utility}. The result was presented as evidence that noisy, pre-fault-tolerant hardware could already provide computational value for specific problem instances. It is worth noting, in the interest of giving a complete picture, that this claim did not go unchallenged: within weeks, improved classical tensor-network and Pauli-path simulation methods reproduced the same results on conventional hardware, in some cases faster than the quantum experiment itself \cite{tindall2024,begusic2023}. The episode is a useful illustration of the current state of the field: the boundary between problems only a QPU can solve and problems classical algorithms can still catch up to is actively contested and moves with each new classical algorithmic technique, which is precisely why the fault-tolerant, error-corrected results discussed in Section~\ref{subsec:qec} matter, they target a regime where classical simulation is provably, not just currently, infeasible.

\subsection{Material Innovations}
\label{subsec:materials}

Material choices directly determine how much of the coherence budget discussed in Section~\ref{subsec:decoherence-mitigation} is available for computation. Quasiparticle tunneling across the junction remains a primary dissipative mechanism, and the superconducting energy gap of the chosen material governs how strongly the junction is protected against it; material and fabrication choices that reduce quasiparticles trapped near the junction directly reduce this channel \cite{MartinisOsborne}. Several material-level threads are active:

\begin{itemize}
    \item \textbf{Aluminum vs.\ niobium.} Aluminum-based junctions have historically shown longer coherence than niobium, plausibly because aluminum's superconducting gap increases favorably with oxygen incorporation and other scattering defects \cite{MartinisOsborne}. However, niobium has seen a genuine resurgence: recent fabrication advances in optically defined niobium trilayer junctions report coherence competitive with aluminum while enabling higher operating temperatures and frequencies, reopening niobium as a viable choice where its other advantages (higher critical temperature, established fabrication infrastructure) are valuable \cite{anferov2024}.
    \item \textbf{Fabrication reproducibility.} As discussed in Section~\ref{subsec:decoherence-mitigation}, junction-to-junction variation from the fabrication process is itself an effective noise source at the chip level; techniques such as optimized electron-beam lithography dose control and the use of high-purity trilayer films (e.g., Al/Nb/Al on silicon) improve yield and consistency \cite{Rebello2024}.
    \item \textbf{Dielectric and surface loss.} Capacitive loss from sub-micron junction fabrication, and the quality of superconducting film surfaces and interfaces, including grain size, grain-boundary grooves, and oxide barrier thickness uniformity in Al/AlO$_x$/Al junctions, measurably affect coherence and remain active areas of characterization \cite{Rebello2024}.
    \item \textbf{Post-fabrication processing.} Laser annealing of transmon qubits after fabrication has been demonstrated as a technique for correcting frequency targeting errors and improving device yield on high-fidelity superconducting processors, without requiring a full refabrication cycle \cite{Zhang2020,Kim2022}.
\end{itemize}

\subsection{Advances in Josephson Junction Physics}

Beyond materials, several directions in junction physics itself are actively reshaping qubit design:

\begin{itemize}
    \item \textbf{Alternative sources of nonlinearity.} Anharmonicity is what makes a qubit's lowest two levels individually addressable (Section~\ref{subsec:junction-to-qubit}); phase-slip junctions have been explored as an alternative source of nonlinearity to the conventional Josephson junction, offering potentially higher operating frequencies and temperatures and pointing toward new classes of intrinsically protected qubits \cite{Purmessur2025}.
    \item \textbf{Quasiparticle dynamics and gap engineering.} Deliberately engineering different superconducting gaps across a junction, rather than using a uniform material on both sides, has been shown to suppress quasiparticle tunneling and resist the resulting errors, an approach directly targeting the dominant decoherence channel identified in Section~\ref{subsec:decoherence-mitigation} \cite{Kurter2022}.
\end{itemize}

\subsection{Hybrid Systems}

No single qubit modality is simultaneously the best at computation, memory, and long-distance communication, which motivates hybrid architectures that combine complementary strengths. Photons interact only weakly with their environment, making them well suited to carrying quantum information over distance, but this same weak interaction makes two-qubit gates between photons difficult (Section~\ref{sec:comparison}); stationary qubits such as superconducting circuits or trapped ions are the reverse, well suited to computation but harder to move \cite{NielsenChuang}. This complementarity motivates a division of labor: stationary qubits handle computation, while "flying" qubits, typically photons, handle communication between modules, connected via quantum teleportation over purified entangled pairs \cite{NielsenChuang}. Molecular ensembles have also been proposed as quantum memory elements for solid-state circuits, illustrating the broader interest in combining disparate quantum systems to compensate for one another's weaknesses \cite{NielsenChuang}. Whether such hybrid approaches become a practical scaling strategy for superconducting QPUs specifically, or remain confined to specialized communication links, is likely to be decided by how the classical-interconnect and error-correction overheads discussed in Section~\ref{sec:challenges} evolve over the next several years.

With this picture of applications, materials, and near-term results in place, we now compare Josephson-junction QPUs directly against the other major hardware modalities pursuing the same computational goals.

\section{Alternative QPU Architectures and Comparative Analysis}
\label{sec:comparison}

Superconducting circuits are one of several serious approaches to building a QPU, and none of the alternatives has yet been eliminated from contention. This section reviews the three most developed alternatives, trapped ions, photons, and neutral atoms, and closes with a direct
comparison to the Josephson-junction platform described in Sections~\ref{sec:fundamentals}-\ref{sec:applications}. Each modality makes a different bet about which physical property is easiest to protect and which is acceptable to sacrifice.

\subsection{Ion Trap QPUs}

Ion trap QPUs confine individual atomic ions with electromagnetic fields, typically in a linear Paul trap using a combination of RF and static fields to create a pseudopotential well \cite{BruzewiczChiaveriniMcConnellSage,BrownChiaveriniSageHaffner}. Quantum information is stored in long-lived internal electronic states, commonly hyperfine states separated by GHz frequencies, and manipulated with laser or microwave fields \cite{BruzewiczChiaveriniMcConnellSage}. Single-qubit gates act on individual ions via resonant radiation; two-qubit gates typically use the ions' shared vibrational motion as a quantum bus, from the original Cirac-Zoller proposal to the more widely used M{\o}lmer-S{\o}rensen gate, which avoids the need to address single ions individually \cite{BruzewiczChiaveriniMcConnellSage,Steane2001}.

Ion traps offer the highest reported gate fidelities of any qubit modality, single-qubit fidelities near 99.9999\% and two-qubit fidelities around 99.9\% in leading experiments, alongside coherence times reaching seconds to minutes, since ions are naturally identical and well isolated from their environment \cite{BruzewiczChiaveriniMcConnellSage}. Their central weaknesses are speed and scale: two-qubit gates take tens to hundreds of microseconds, orders of magnitude slower than microwave-driven superconducting gates, and while architectures such as the Quantum Charge-Coupled Device (QCCD) address scalability by shuttling ions between dedicated storage, interaction, and readout zones, managing large ion chains, motional heating during transport, and the laser and control infrastructure needed to individually address many ions remain significant open engineering problems \cite{BruzewiczChiaveriniMcConnellSage,Kielpinski2002}.

\subsection{Photonic QPUs}

Photonic QPUs encode information in single photons, typically via polarization, spatial path, or arrival time, and manipulate it with linear optical elements such as beam splitters and phase shifters, with photon detection providing both readout and, in some schemes, the gate mechanism itself \cite{Pieter2007}. Single-qubit gates are straightforward and deterministic; two-qubit entangling gates are the platform's defining difficulty, since photons do not naturally interact. The Knill-Laflamme-Milburn protocol showed that scalable computation is possible anyway, using probabilistic gates based on photon interference (the Hong-Ou-Mandel effect) combined with heralding and error correction \cite{Pieter2007}; more recent approaches include cluster-state computing, where an entangled multi-photon resource state is prepared offline and computation proceeds via single-qubit measurements, and time-domain multiplexing, which reuses a small number of optical components to process many sequential pulses \cite{Pieter2007,Madsen2022}.

Photons interact only weakly with their environment, giving photonic QPUs low decoherence and making them the natural carrier for quantum communication over long distances \cite{Pieter2007}. Photonic processors have also demonstrated a genuine computational advantage on a specific sampling task, Gaussian Boson Sampling, outperforming classical supercomputers on that narrow benchmark \cite{Madsen2022}. The same weak photon-photon
interaction that gives photonics its coherence advantage is also its central liability: deterministic two-qubit gates require strong optical nonlinearities that are inherently weak at the single-photon level, so probabilistic schemes carry a substantial resource overhead, and building large photonic processors requires high-efficiency, well-matched sources,
low-loss circuits, and photon-number-resolving detectors simultaneously \cite{Pieter2007}. Photon loss itself is a distinct and persistent error channel, and dedicated loss-tolerant error correction, rather than the generic codes discussed in Section~\ref{subsec:qec}, is generally needed to address it \cite{Pieter2007}.

\subsection{Neutral Atom QPUs}
\label{subsec:neutral-atom}

Neutral atom QPUs trap individual atoms, typically rubidium or strontium, in arrays of tightly focused laser beams called optical tweezers, with quantum information stored in long-lived hyperfine ground states \cite{saffman2016,henriet2020}. Two-qubit gates exploit the \emph{Rydberg blockade}: exciting an atom to a highly excited Rydberg state shifts the energy levels of its neighbors strongly enough to prevent them from being excited to the same state simultaneously, and this conditional behavior is used to implement entangling gates \cite{saffman2016}.

Two properties distinguish this platform from the others surveyed here. First, tweezer-trapped atoms are \emph{room-temperature} qubits: the vacuum chamber and optics operate without a dilution refrigerator, in contrast to every Josephson-junction QPU discussed in this paper \cite{henriet2020}. Second, and more consequentially for scaling, optical tweezers give neutral-atom arrays \emph{reconfigurable connectivity}, atoms can be physically shuttled mid-circuit to bring new pairs into interaction range, rather than being restricted to fixed nearest-neighbor coupling as on a superconducting chip \cite{saffman2016}. This flexibility directly enables more efficient error-correction layouts, since a logical qubit's constituent atoms are not confined to a fixed local patch. Gate times fall between the other two established platforms: hundreds of nanoseconds to a few microseconds for Rydberg gates, faster than trapped-ion gates but slower than superconducting CZ gates (20-100~ns) \cite{postquantumneutral}. As of early 2026, leading two-qubit Rydberg gate fidelities are around 99.5\%, still below the best reported trapped-ion fidelities, and closing this gap is the platform's main outstanding technical challenge \cite{postquantumneutral}.

Neutral atoms have moved from a promising proposal to a demonstrated fault-tolerant testbed on a timescale directly comparable to the superconducting result discussed in Section~\ref{subsec:qec}: using reconfigurable arrays of up to 448 atoms segmented into storage, entangling, readout, and reservoir zones, a below-threshold surface-code memory was demonstrated with a $2.14(13)\times$ suppression in logical error per added code distance, essentially the same suppression factor Google reported for Willow, together with a demonstration of universal logical gates via transversal operations and lattice surgery \cite{bluvstein2026}. The architecture's scalability case rests on optical engineering rather than new physics: extending the array primarily requires projecting more tweezers and refining atom-loss detection and reloading, rather than solving a new class of fabrication problem \cite{postquantumneutral}. Atom loss during computation, an atom escaping its trap mid-circuit, is the platform's most distinctive error channel, and mitigating it, alongside closing the gate-fidelity gap to trapped ions, are its central open engineering problems.

\subsection{Comparative Summary}

Table~\ref{tab:comparison} summarizes the four architectures across the metrics most relevant to building a fault-tolerant QPU.

\begin{table*}[t]
\centering
\small
\renewcommand{\arraystretch}{1.35}
\rowcolors{2}{gray!12}{white}
\begin{tabular}{
  >{\raggedright\arraybackslash}m{2.1cm}
  >{\raggedright\arraybackslash}m{2.6cm}
  >{\raggedright\arraybackslash}m{2.6cm}
  >{\raggedright\arraybackslash}m{2.6cm}
  >{\raggedright\arraybackslash}m{2.6cm}
}
\toprule
\rowcolor{white}
\textbf{Feature} & \textbf{Josephson Junction Superconducting} & \textbf{Ion Trap} & \textbf{Photonic} & \textbf{Neutral Atom} \\
\midrule
Coherence time & Relatively short; mitigated by materials and QEC & Longest (s to minutes) & Long (weak environmental coupling) & Long; room-temperature operation \\
Gate fidelity & High (optimal-control pulses) & Highest (1-qubit $\sim$99.9999\%, 2-qubit $\sim$99.9\%) & Variable; entangling gates remain hard & High and improving ($\sim$99.5\% 2-qubit) \\
Gate speed & Fast (20-100~ns, microwave) & Slower (tens-hundreds of $\mu$s) & Fast in principle; deterministic gates slow & Intermediate (hundreds of ns-few $\mu$s) \\
Connectivity & Limited, fixed nearest-neighbor & High, via shared motion & Network-focused & Reconfigurable via tweezer shuttling \\
Operating temp. & Ultra-low (mK, dilution fridge) & Vacuum + laser cooling & Room temperature & Room temperature \\
Noise sensitivity & Sensitive; requires shielding and mK cooling & Low, but motional heating is a challenge & Low, but photon loss is a persistent error channel & Low decoherence; atom loss during computation is the distinctive channel \\
Scalability outlook & Chip-based, but crosstalk and control I/O limit scale & Challenging (chain size, motional heating) & Challenging (probabilistic gates, component integration) & Promising; primarily an optical-engineering problem \\
Applications & General-purpose computing, cryptography, quantum simulation & High-fidelity computing, simulation & Quantum communication, networking & Analog many-body simulation, combinatorial optimization, increasingly general-purpose computing \\
Demonstrated FTQC milestone & Below-threshold surface code, Willow, 2025 \cite{acharya2025willow} & - & - & Below-threshold surface code + universal logic, 2026 \cite{bluvstein2026} \\
\bottomrule
\end{tabular}
\caption{Comparison of Quantum Processing Unit technologies.}
\label{tab:comparison}
\end{table*}

Superconducting and neutral-atom platforms are, as of this writing, the only two modalities with a published below-threshold fault-tolerance demonstration, achieved within roughly a year of one another and at a similar logical-error suppression factor per code distance \cite{acharya2025willow,bluvstein2026}. This convergence is notable: it suggests the surface code's threshold behavior is a robust target reachable by more than one physical platform, which reframes the competition between architectures. The open question is no longer only whether a platform \emph{can} cross the fault-tolerance threshold, but which one can do so while also solving the classical control-interface problem (Section~\ref{subsec:control-complexity}) at the qubit counts fault-tolerant algorithms will actually require. Superconducting QPUs retain a clear speed advantage and the most mature chip-fabrication supply chain; neutral atoms offer reconfigurable connectivity and room-temperature operation that meaningfully ease the scaling problem discussed throughout Section~\ref{sec:challenges}; ion traps remain the fidelity benchmark other platforms are measured against; and photonic systems retain a distinct role in quantum communication regardless of how the computational comparison resolves. Each platform's roadmap over the next several years will likely be judged less by any single benchmark and more by how directly it can turn a below-threshold logical qubit into a large number of them.

\section{Outlook: The Field Since Early 2025}
\label{sec:outlook}

The results discussed in Sections~\ref{sec:challenges}-\ref{sec:comparison}, below-threshold surface-code memories on both superconducting and neutral-atom hardware, continued materials progress, and a maturing comparative landscape, raise an obvious question: what happens next,
and how far is "next" from a computer that solves problems of practical interest? This section lays out the two most detailed public roadmaps for superconducting QPUs specifically, then calibrates the gap that remains.

\subsection{Two Roadmaps to Fault Tolerance}

IBM and Google have converged on the same target, a large-scale, error-corrected superconducting quantum computer by roughly the end of the decade, via different intermediate strategies.

\textbf{IBM} has committed to a modular, code-first roadmap. Its 2024 demonstration of high-rate \emph{bivariate bicycle} qLDPC codes (Section~\ref{subsec:qec}) showed that qLDPC encoding can cut physical qubit overhead relative to the surface code by roughly an order of
magnitude \cite{bravyi2024,ibmroadmap2025}. The roadmap sequences the hardware needed to exploit this: Loon (2025) introduces \emph{c-couplers} providing longer-range on-chip connectivity, a prerequisite for qLDPC codes that the nearest-neighbor connectivity discussed in Section~\ref{subsec:junction-to-qubit} cannot support; Kookaburra (2026) is intended as the first module combining qLDPC quantum memory with a logical processing unit; Cockatoo (2027) entangles two such modules; and Starling, targeted for 2028-2029, is intended to combine multiple modules into a 200-logical-qubit, 100-million-gate fault-tolerant system, with a successor (Blue Jay, targeted for 2033 and beyond) intended to reach roughly 2{,}000 logical qubits \cite{ibmroadmap2025}. IBM has also separately targeted a demonstration of quantum advantage on interim, pre-fault-tolerant hardware by the end of 2026 \cite{ibmroadmap2025}.

\textbf{Google} frames its progress as a six-milestone sequence, with Willow's below-threshold result (Section~\ref{subsec:qec}) representing roughly the second milestone, on the order of 100 physical qubits with a logical error rate near $10^{-2}$, and a long-lived logical qubit and a logical two-qubit gate between separate logical qubits as the next targets \cite{googleroadmap2026}. In October 2025, Google additionally reported what it characterized as a verifiable quantum advantage on Willow using an algorithm based on out-of-time-order correlators, dubbed Quantum Echoes, reporting a roughly 13{,}000-fold speedup over the best available classical estimate for a chaotic-dynamics benchmark with a result independently checkable by comparison against nuclear magnetic resonance measurements \cite{googleqa2025echoes}. In March 2026, Google extended its strategy to a second, complementary hardware modality, adding neutral-atom development (Section~\ref{subsec:neutral-atom}) alongside its superconducting program, reasoning that superconducting qubits scale well in circuit depth while neutral-atom arrays scale well in raw qubit count \cite{googleroadmap2026}.

Other efforts are pursuing variations on the same underlying bet: reduce the physical-to-logical qubit ratio faster than qubit counts grow. Notably, dual-rail qubit encodings, which convert the dominant superconducting error channel, photon loss, into a detectable erasure rather than an undetected bit-flip, have been proposed as a route to higher-quality logical qubits with a smaller decoding burden than the standard transmon-based surface code \cite{dwaveroadmap2026}.

\subsection{Calibrating the Remaining Gap}

It is worth stating plainly how much distance separates a below-threshold demonstration from a computer capable of running an algorithm like Shor's factoring algorithm (Section~\ref{sec:introduction}) at cryptographically relevant scale. The distance-7 logical error rate reported for Willow, approximately $1.4\times10^{-3}$ per cycle, is itself a landmark, but factoring a 2048-bit RSA key is estimated to require logical error rates on the order of $10^{-10}$ or better sustained over the full computation, a gap of roughly seven orders of magnitude beyond what has been demonstrated to date \cite{acharya2025willow}. Closing that gap is explicitly the purpose of the multi-year roadmaps above: qLDPC codes targeting an order-of-magnitude reduction in overhead \cite{bravyi2024,ibmroadmap2025}, continued materials and fabrication work targeting lower baseline physical error rates (Section~\ref{subsec:materials}), and classical control-interface scaling (Section~\ref{subsec:control-complexity}) all attack different parts of the same problem simultaneously. None of this diminishes the significance of a below-threshold result, it is the first experimental confirmation that the threshold theorem's central premise holds on real hardware, but the honest framing is that the field has demonstrated the \emph{sign} of the effect it needs, not yet the \emph{magnitude}.

\subsection{Open Problems}

Three questions currently determine how quickly the roadmaps above can be realized, and none is settled:

\begin{itemize}
    \item \textbf{Code choice.} Whether qLDPC codes' overhead advantage outweighs the fabrication complexity of the longer-range couplers they require, relative to a more conservative surface-code path, remains an open systems-engineering bet rather than a solved question (Section~\ref{subsec:qec}).
    \item \textbf{Classical I/O at scale.} Every roadmap above assumes cryo-CMOS and multiplexing solutions (Section~\ref{subsec:control-complexity}) continue to scale roughly in step with physical qubit count; this has not yet been demonstrated at the qubit counts Starling-class systems will require.
    \item \textbf{Cross-modality competition.} With superconducting and neutral-atom platforms now both past the below-threshold milestone within roughly a year of one another (Section~\ref{subsec:neutral-atom}), and with dual-rail and trapped-ion efforts pursuing their own paths to lower overhead, it is not yet clear which platform, or combination of platforms, will reach useful scale first.
\end{itemize}

These are engineering and systems questions more than open physics questions, which is itself a marker of how far the field has moved since this paper's first version. We return to this framing in the conclusion.

\section{Conclusion}
\label{sec:conclusion}

This paper has traced Josephson-junction QPUs from first principles to current practice: the quantum-mechanical foundations that any qubit must respect, the specific way a shunted junction becomes an addressable two-level system, and the engineering effort required to keep thousands of such systems coherent, controllable, and correctable at once. Two conclusions follow from this account. First, the central obstacles facing superconducting QPUs are no longer purely questions of physics. Materials and circuit design have advanced to the point that the threshold theorem's central prediction, that error-corrected logical qubits can outperform their physical constituents, has now been demonstrated experimentally, on superconducting hardware and, within roughly a year, on neutral-atom hardware as well. What remains is substantially a systems-engineering problem: reducing physical-to-logical qubit overhead through codes such as qLDPC, scaling classical control electronics without a proportional increase in cost and heat load, and closing an error-rate gap of several more orders of magnitude before cryptographically or scientifically consequential algorithms become practical. Second, no single hardware modality has settled the competition. Superconducting circuits retain the fastest gates and the most mature fabrication base; trapped ions retain the highest fidelities; photonic systems retain an essentially unique role in quantum communication; and neutral atoms have emerged as a credible fourth path, sufficiently so that organizations previously committed to a single modality are now pursuing more than one in parallel. Given this, the near-term trajectory of the field is likely to be set less by which platform is fundamentally superior and more by which combination of platforms, codes, and control architectures first converts a below-threshold laboratory result into a machine that runs an algorithm no classical computer can match. Josephson junctions, a decade into their role as the leading qubit technology, remain central to that outcome.

\section*{Acknowledgment}
We would like to express our sincere gratitude to Dr. Said Jawad Saidi for his valuable feedback and insightful comments during the writing of this paper. His guidance has been instrumental in improving the quality of our work. Additionally, I acknowledge the use of ChatGPT for grammar correction and text clarity enhancement, which helped refine the language and coherence of this paper.

\bibliographystyle{plain} 
\bibliography{ref.bib}

@article{Shor,
  author  = {Shor, Peter W.},
  title   = {Polynomial-time algorithms for prime factorization and discrete logarithms on a quantum computer},
  journal = {SIAM Journal on Computing},
  year    = {1997},
}

@book{NielsenChuang,
  author    = {Nielsen, Michael A. and Chuang, Isaac L.},
  title     = {Quantum Computation and Quantum Information},
  year      = {2000},
  publisher = {Cambridge University Press},
}

@article{CaoRomeroAspuru,
  author  = {Cao, Yudong and Romero, Jonathan and Aspuru-Guzik, Alán},
  title   = {Potential of quantum computing for drug discovery},
  journal = {Nature Reviews Drug Discovery},
  year    = {2019},
}

@article{SchuldSinayskiyPetruccione,
  author  = {Schuld, Maria and Sinayskiy, Ilya and Petruccione, Francesco},
  title   = {An introduction to quantum machine learning},
  journal = {Contemporary Physics},
  year    = {2015},
}

@article{FarhiGoldstoneGutmann,
  author  = {Farhi, Edward and Goldstone, Jeffrey and Gutmann, Sam},
  title   = {A quantum approximate optimization algorithm},
  journal = {arXiv preprint},
  year    = {2014},
  eprint  = {arXiv:1411.4028},
}

@article{acharya2025willow,
  title     = {Quantum error correction below the surface code threshold},
  author    = {{Google Quantum AI and Collaborators} and Acharya, Rajeev and Aghababaie-Beni, Laleh and Aleiner, Igor and Andersen, Trond I. and Ansmann, Markus and Arute, Frank and Arya, Kunal and Asfaw, Abraham and Astrakhantsev, Nikita and others},
  journal   = {Nature},
  volume    = {638},
  number    = {8052},
  pages     = {920--926},
  year      = {2025},
  publisher = {Nature Publishing Group},
  doi       = {10.1038/s41586-024-08449-y}
}

@phdthesis{TurroThesis,
  author       = {Turro, F.},
  title        = {Quantum Algorithms for Many-Body Structure and Dynamics},
  year         = {2022},
  school = {University of Trento},
  supervisor   = {Prof. Francesco Pederiva},
  cosupervisor = {Dr. Iacopo Carusotto},
  note = {Physics Department. Defense date: 10 June 2022},
}

@inproceedings{Grover1996,
  author    = {Grover, Lov K.},
  title     = {A Fast Quantum Mechanical Algorithm for Database Search},
  booktitle = {Proceedings of the Twenty-Eighth Annual ACM Symposium on Theory of Computing (STOC '96)},
  year      = {1996},
  pages     = {212--219},
  publisher = {Association for Computing Machinery},
  address   = {Philadelphia, Pennsylvania, USA},
  isbn      = {0897917855},
  doi       = {10.1145/237814.237866},
  url       = {https://doi.org/10.1145/237814.237866},
}

@inproceedings{Shor1994,
  author    = {Shor, Peter W.},
  title     = {Algorithms for quantum computation: discrete logarithms and factoring},
  booktitle = {Proceedings 35th Annual Symposium on Foundations of Computer Science},
  year      = {1994},
  pages     = {124--134},
}

@article{Clauser1969,
  author = {Clauser, J. F. and Horne, M. A. and Shimony, A. and Holt, R. A.},
  title   = {Proposed Experiment to Test Local Hidden-Variable Theories},
  journal = {Phys. Rev. Lett.},
  volume  = {23},
  number  = {15},
  pages   = {880--884},
  year    = {1969},
  doi     = {10.1103/PhysRevLett.23.880},
  url     = {https://link.aps.org/doi/10.1103/PhysRevLett.23.880},
}

@article{Bell1964,
  author  = {Bell, J. S.},
  title   = {On the Einstein Podolsky Rosen paradox},
  journal = {Phys. Phys. Fiz.},
  volume  = {1},
  number  = {3},
  pages   = {195--200},
  year    = {1964},
  doi     = {10.1103/PhysicsPhysiqueFizika.1.195},
  url     = {https://link.aps.org/doi/10.1103/PhysicsPhysiqueFizika.1.195},
}

@article{EinsteinPodolskyRosen1935,
  author  = {Einstein, A. and Podolsky, B. and Rosen, N.},
  title   = {Can Quantum-Mechanical Description of Physical Reality Be Considered Complete?},
  journal = {Phys. Rev.},
  volume  = {47},
  number  = {10},
  pages   = {777--780},
  year    = {1935},
  doi     = {10.1103/PhysRev.47.777},
  url     = {https://link.aps.org/doi/10.1103/PhysRev.47.777},
}

@article{Catelani2012,
  author  = {Catelani, G. and others},
  title   = {Decoherence of superconducting qubits caused by quasiparticle tunneling},
  journal = {Physical Review B},
  volume  = {86},
  number  = {18},
  year    = {2012},
  doi     = {10.1103/physrevb.86.184514},
  url     = {https://doi.org/10.1103\%2Fphysrevb.86.184514},
}

@techreport{MartinisOsborne,
  author       = {Martinis, John M. and Osborne, K.},
  title        = {Superconducting Qubits and the Physics of Josephson Junctions},
  institution = {National Institute of Standards and Technology},
  address = {Boulder, CO, USA},
}

@article{BarencoEtAl,
  author  = {Barenco, A. and others},
  title   = {Elementary gates for quantum computation},
  journal = {Phys. Rev. A},
  volume  = {52},
  number  = {5},
  pages   = {3457--3467},
  year    = {1995},
  issn    = {1094-1622},
  doi     = {10.1103/physreva.52.3457},
  url     = {http://dx.doi.org/10.1103/PhysRevA.52.3457},
}

@article{VatanWilliams,
  author  = {Vatan, F. and Williams, C.},
  title   = {Optimal quantum circuits for general two-qubit gates},
  journal = {Physical Review A: Atomic, Molecular, and Optical Physics},
  volume  = {69},
  number  = {3},
  year    = {2004},
  issn    = {1094-1622},
  doi     = {10.1103/physreva.69.032315},
  url     = {http://dx.doi.org/10.1103/PhysRevA.69.032315},
}

@book{ClarkeBraginski,
  editor    = {Clarke, John and Braginski, Alex I.},
  title     = {The SQUID Handbook},
  year      = {2006},
  publisher = {Wiley-VCH},
}

@book{Tinkham,
  author    = {Tinkham, Michael},
  title     = {Introduction to Superconductivity},
  year      = {2004},
  publisher = {Dover Publications},
}

@misc{Suzuki2012,
  author       = {Suzuki, M. S.},
  title        = {The Josephson Effect},
  year         = {2012},
  howpublished = {\url{https://bingweb.binghamton.edu/~suzuki/ModernPhysics/35_Josephson_effect.pdf}},
  note = {Lecture notes, Department of Physics, State University of New York at Binghamton.},
}

@article{Krantz2019,
  author    = {Krantz, P. and others},
  title     = {A quantum engineer’s guide to superconducting qubits},
  journal   = {Applied Physics Reviews},
  year      = {2019},
  issn      = {1931-9401},
  doi       = {10.1063/1.5089550},
  url       = {http://dx.doi.org/10.1063/1.5089550},
}

@article{manucharyan2009,
  title   = {Fluxonium: Single {C}ooper-Pair Circuit Free of Charge Offsets},
  author  = {Manucharyan, Vladimir E. and Koch, Jens and Glazman, Leonid I. and Devoret, Michel H.},
  journal = {Science},
  volume  = {326},
  number  = {5949},
  pages   = {113--116},
  year    = {2009},
  doi     = {10.1126/science.1175552}
}

@article{nguyen2019,
  title   = {High-Coherence Fluxonium Qubit},
  author  = {Nguyen, Long B. and Minev, Zlatko K. and Vool, Uri and Devoret, Michel H. and others},
  journal = {Physical Review X},
  volume  = {9},
  pages   = {041041},
  year    = {2019},
  doi     = {10.1103/PhysRevX.9.041041}
}

@phdthesis{Rebello2024,
  author       = {Rebello, A. C. M.},
  title        = {Advanced Lithography Methods for Creating Josephson Junctions and Superconducting Circuits at Nano and Micro Scales},
school = {Brazilian Center of Physics Research},
year = {2024},
}

@inproceedings{fowler2012,
  title     = {Surface codes: Towards practical large-scale quantum computation},
  author    = {Fowler, Austin G. and Mariantoni, Matteo and Martinis, John M. and Cleland, Andrew N.},
  journal   = {Physical Review A},
  volume    = {86},
  pages     = {032324},
  year      = {2012},
  doi       = {10.1103/PhysRevA.86.032324}
}

@article{bravyi2024,
  title   = {High-threshold and low-overhead fault-tolerant quantum memory},
  author  = {Bravyi, Sergey and Cross, Andrew W. and Gambetta, Jay M. and Maslov, Dmitri and Rall, Patrick and Yoder, Theodore J.},
  journal = {Nature},
  volume  = {627},
  pages   = {778--782},
  year    = {2024},
  doi     = {10.1038/s41586-024-07107-7}
}

@article{patra2018,
  title   = {Cryo-{CMOS} Circuits and Systems for Quantum Computing Applications},
  author  = {Patra, Bishnu and Incandela, Rosario M. and van Dijk, Jeroen P. G. and Homulle, Harald A. R. and others},
  journal = {IEEE Journal of Solid-State Circuits},
  volume  = {53},
  number  = {1},
  pages   = {309--321},
  year    = {2018},
  doi     = {10.1109/JSSC.2017.2737549}
}

@article{Jolin2020,
  author    = {Jolin, S. W. and others},
  title     = {Calibration of mixer amplitude and phase imbalance in superconducting circuits},
  journal   = {Review of Scientific Instruments},
  year      = {2020},
  issn      = {1089-7623},
  doi       = {10.1063/5.0025836},
  url       = {http://dx.doi.org/10.1063/5.0025836},
}

@article{Ayala2023,
  author    = {Ayala, Christopher L. and Tanaka, Tomoyuki and Saito, Ro and Nozoe, Mai and Takeuchi, Naoki and Yoshikawa, Nobuyuki},
  title     = {MANA: A Monolithic Adiabatic iNtegration Architecture Microprocessor Using 1.4-zJ/op Unshunted Superconductor Josephson Junction Devices},
  journal   = {IEEE Transactions on Applied Superconductivity},
  year      = {2021}

}

@article{kim2023utility,
  title   = {Evidence for the utility of quantum computing before fault tolerance},
  author  = {Kim, Youngseok and Eddins, Andrew and Anand, Sajant and Wei, Ken Xuan and van den Berg, Ewout and Rosenblatt, Sami and Nayfeh, Hasan and Wu, Yantao and Zaletel, Michael and Temme, Kristan and Kandala, Abhinav},
  journal = {Nature},
  volume  = {618},
  pages   = {500--505},
  year    = {2023},
  doi     = {10.1038/s41586-023-06096-3}
}

@article{tindall2024,
  title   = {Efficient Tensor Network Simulation of {IBM}'s Eagle Kicked Ising Experiment},
  author  = {Tindall, Joseph and Fishman, Matthew and Stoudenmire, E. Miles and Sels, Dries},
  journal = {PRX Quantum},
  volume  = {5},
  pages   = {010308},
  year    = {2024},
  doi     = {10.1103/PRXQuantum.5.010308}
}

@article{begusic2023,
  title   = {Fast and converged classical simulations of evidence for the utility of quantum computing before fault tolerance},
  author  = {Begu\v{s}i\'c, Tomislav and Chan, Garnet Kin-Lic},
  journal = {Science Advances},
  volume  = {10},
  number  = {3},
  year    = {2024},
  doi     = {10.1126/sciadv.adk4321}
}

@article{anferov2024,
  title   = {Improved Coherence in Optically Defined Niobium Trilayer-Junction Qubits},
  author  = {Anferov, Alexander and Lee, Kan-Heng and Zhao, Fang and Simon, Jonathan and Schuster, David I.},
  journal = {Physical Review Applied},
  volume  = {21},
  pages   = {024047},
  year    = {2024},
  doi     = {10.1103/PhysRevApplied.21.024047}
}

@misc{Zhang2020,
  author  = {Zhang, E. J. and others},
  title   = {High-fidelity superconducting quantum processors via laser-annealing of transmon qubits},
  year    = {2020},
  eprint  = {arXiv:2012.08475},
  archivePrefix = {arXiv},
  primaryClass = {quant-ph},
}

@article{Kurter2022,
  author  = {Kurter, C. and Murray, C. E. and Gordon, R. T. and others},
  title   = {Quasiparticle tunneling as a probe of Josephson junction barrier and capacitor material in superconducting qubits},
  journal = {npj Quantum Information},
  year    = {2022},
  volume  = {8},
  pages   = {31},
  month   = {March},
  doi     = {10.1038/s41534-022-00542-2},
  url     = {https://doi.org/10.1038/s41534-022-00542-2},
  note    = {Published: 21 March 2022}
}

@misc{Purmessur2025,
  author  = {Purmessur, Cheeranjeev and Chow, Kaicheung and van Heck, Bernard and Kou, Angela},
  title   = {Operation of a high-frequency, phase-slip qubit},
  year    = {2025},
  eprint  = {2502.07043},
  archivePrefix = {arXiv},
  primaryClass = {cond-mat.supr-con},
  url     = {https://arxiv.org/abs/2502.07043}
}

@article{BruzewiczChiaveriniMcConnellSage,
  author  = {Bruzewicz, Colin D. and Chiaverini, John and McConnell, Robert and Sage, Jeremy M.},
  title   = {Trapped-ion quantum computing: Progress and challenges},
  journal = {Applied Physics Reviews},
  year    = {2019},
  volume  = {6},
  pages   = {021314},
  doi     = {10.1063/1.5088164},
  url     = {https://doi.org/10.1063/1.5088164}
}

@article{BrownChiaveriniSageHaffner,
  author  = {Brown, Kenneth R. and Chiaverini, John and Sage, Jeremy M. and H{\"a}ffner, Hartmut},
  title   = {Materials challenges for trapped-ion quantum computers},
  journal = {Nature Reviews Materials},
  year    = {2021},
  volume  = {6},
  pages   = {892--905},
  month   = {October},
  doi     = {10.1038/s41578-021-00292-1},
  url     = {https://doi.org/10.1038/s41578-021-00292-1}
}

@inproceedings{Steane2001,
  author    = {Steane, Andrew M.},
  title     = {Quantum computing with trapped ions, atoms and light},
  booktitle = {AIP Conference Proceedings},
  year      = {2001},
  volume    = {551},
  pages     = {158--172},
  month     = {January},
  doi       = {10.1063/1.1354347},
  url       = {https://doi.org/10.1063/1.1354347}
}

@article{Kielpinski2002,
  author  = {Kielpinski, D. and Monroe, C. and Wineland, D.},
  title   = {Architecture for a large-scale ion-trap quantum computer},
  journal = {Nature},
  year    = {2002},
  volume  = {417},
  pages   = {709--711},
  month   = {June},
  day    = {13},
  doi     = {10.1038/nature00784},
  url     = {https://doi.org/10.1038/nature00784}
}

@article{Pieter2007,
  title = {Linear optical quantum computing with photonic qubits},
  author = {Kok, Pieter and Munro, W. J. and Nemoto, Kae and Ralph, T. C. and Dowling, Jonathan P. and Milburn, G. J.},
  journal = {Rev. Mod. Phys.},
  volume = {79},
  issue = {1},
  pages = {135--174},
  numpages = {0},
  year = {2007},
  month = {Jan},
  publisher = {American Physical Society},
  doi = {10.1103/RevModPhys.79.135},
  url = {https://link.aps.org/doi/10.1103/RevModPhys.79.135}
}

@article{Madsen2022,
  author  = {Madsen, Lars S. and Laudenbach, Fabian and Askarani, Mohsen F. and others},
  title   = {Quantum computational advantage with a programmable photonic processor},
  journal = {Nature},
  year    = {2022},
  volume  = {606},
  pages   = {75--81},
  month   = {June},
  day     = {02},
  doi     = {10.1038/s41586-022-04725-x},
  url     = {https://doi.org/10.1038/s41586-022-04725-x},
  note    = {Published: 01 June 2022}
}

@article{bluvstein2026,
  title   = {A fault-tolerant neutral-atom architecture for universal quantum computation},
  author  = {Bluvstein, Dolev and Geim, Alexandra A. and Li, Sophie H. and Evered, Simon J. and Bonilla Ataides, J. Pablo and Baranes, Gefen and Gu, Andi and Manovitz, Tom and Xu, Muqing and Kalinowski, Marcin and Majidy, Shayan and Kokail, Christian and Maskara, Nishad and Trapp, Elias C. and Stewart, Luke M. and Hollerith, Simon and Zhou, Hengyun and Gullans, Michael J. and Yelin, Susanne F. and Greiner, Markus and Vuleti\'{c}, Vladan and Cain, Madelyn and Lukin, Mikhail D.},
  journal = {Nature},
  volume  = {649},
  pages   = {39--46},
  year    = {2026},
  doi     = {10.1038/s41586-025-09848-5}
}

@article{saffman2016,
  title   = {Quantum computing with atomic qubits and Rydberg interactions: progress and challenges},
  author  = {Saffman, M.},
  journal = {Journal of Physics B: Atomic, Molecular and Optical Physics},
  volume  = {49},
  pages   = {202001},
  year    = {2016},
  doi     = {10.1088/0953-4075/49/20/202001}
}

@article{henriet2020,
  title   = {Quantum computing with neutral atoms},
  author  = {Henriet, Lo\"{i}c and B\'{e}guin, Lucas and Signoles, Adrien and Lahaye, Thierry and Browaeys, Antoine and Reymond, Georges-Olivier and Jurczak, Christophe},
  journal = {Quantum},
  volume  = {4},
  pages   = {327},
  year    = {2020},
  doi     = {10.22331/q-2020-09-21-327}
}

@misc{postquantumneutral,
  title        = {Quantum Computing Modalities: Neutral Atom (Rydberg)},
  howpublished = {\url{https://postquantum.com/quantum-modalities/neutral-atom-quantum/}},
  note         = {Accessed July 2026},
  year         = {2026}
}

@misc{ibmroadmap2025,
  title        = {IBM Lays Out Clear Path to Fault-Tolerant Quantum Computing},
  author       = {{IBM Quantum}},
  howpublished = {\url{https://www.ibm.com/quantum/blog/large-scale-ftqc}},
  note         = {Accessed July 2026},
  year         = {2025}
}

@misc{googleroadmap2026,
  title        = {Google Quantum AI Roadmap: Willow, Neutral Atoms, and the Path to Utility},
  author       = {{Google Quantum AI}},
  howpublished = {\url{https://quantumzeitgeist.com/google-quantum-computing/}},
  note         = {Secondary summary of Google's public roadmap statements; accessed July 2026},
  year         = {2026}
}

@article{googleqa2025echoes,
  title   = {Observation of constructive interference at the edge of quantum ergodicity},
  author  = {{Google Quantum AI and Collaborators}},
  journal = {Nature},
  year    = {2025},
  doi     = {10.1038/s41586-025-09526-6}
}

@misc{dwaveroadmap2026,
  title        = {D-Wave Charts a New Course to Fault-Tolerant Quantum Computing with Gate-Model Roadmap},
  author       = {{D-Wave Quantum Inc.}},
  howpublished = {\url{https://www.sec.gov/Archives/edgar/data/1907982/000190798226000012/d-wave_pressconferencexp.htm}},
  note         = {Accessed July 2026},
  year         = {2026}
}

@article{Kim2022,
  author  = {Kim, H. and others},
  title   = {Effects of laser-annealing on fixed-frequency superconducting qubits},
  journal = {Applied Physics Letters},
  volume  = {121},
  number  = {14},
  pages   = {142601},
  month   = {10},
  year    = {2022},
  doi     = {10.1063/5.0102092},
  url     = {https://doi.org/10.1063/5.0102092},
}

@article{koch2007,
  title   = {Charge-insensitive qubit design derived from the {C}ooper pair box},
  author  = {Koch, Jens and Yu, Terri M. and Gambetta, Jay and Houck, A. A. and Schuster, D. I. and Majer, J. and Blais, Alexandre and Devoret, M. H. and Girvin, S. M. and Schoelkopf, R. J.},
  journal = {Physical Review A},
  volume  = {76},
  pages   = {042319},
  year    = {2007},
  doi     = {10.1103/PhysRevA.76.042319}
}
\end{document}